\documentclass[conference]{IEEEtran}
\usepackage{cite}
\usepackage{amsmath,amssymb}
\usepackage{graphicx}
\usepackage{url}
\usepackage{hyperref}

\title{Identity Control Plane: The Unifying Layer for Zero Trust Infrastructure}

\author{
  \IEEEauthorblockN{Surya Teja Avirneni}
  \IEEEauthorblockA{
    Cloud Platform and Security Architect, \\
    Member: IEEE, ACM, ISC2
  }
}

\begin{document}

\maketitle

\begin{abstract}
Identity has emerged as the foundational control vector in modern Zero Trust infrastructure. However, identity is often fragmented across human, workload, and automation domains, leading to inconsistent policy enforcement, static privileges, and limited auditability. This paper introduces the Identity Control Plane (ICP), a unifying architectural pattern that integrates SPIFFE-based workload identity, OIDC/SAML user identity, and scoped automation tokens. ICP enables dynamic, intent-aware access control using attribute-based policy engines. We describe its core components, modes of enforcement, use cases, and integration patterns. We also include a comparative analysis with existing models and theorize performance characteristics such as token evaluation latency and enforcement complexity. The ICP aligns with IETF WIMSE standards and provides a scalable framework for secure, identity-aware infrastructure orchestration.
\end{abstract}

\section{Introduction}
Zero Trust architectures prioritize identity-aware access control across distributed, multi-tenant systems. As organizations adopt internal developer platforms (IDPs), CI/CD pipelines, and multi-cloud infrastructure, the surface area of identity increases significantly.

Today, identity is fragmented: human users authenticate through OIDC or SAML, workloads may use SPIFFE or IAM roles, and automation often relies on static secrets or long-lived service accounts. This results in inconsistent policy enforcement, over-permissioning, and poor auditability.

This paper introduces the \textbf{Identity Control Plane (ICP)}—a unifying control layer that abstracts identity across domains and enforces fine-grained, dynamic policies based on actor attributes, runtime metadata, and organizational intent.

The ICP model integrates:
\begin{itemize}
    \item SPIFFE-based workload identity for service-to-service authentication~\cite{spiffe},
    \item OIDC/SAML-based identity for human actors,
    \item Broker-issued transaction tokens for scoped automation credentials,
    \item Attribute-based access control (ABAC) enforcement engines such as OPA~\cite{opa} and Cedar~\cite{cedar}.
\end{itemize}
This work builds on concepts from SPIFFE, the OAuth transaction token model, and the IETF WIMSE initiative, which standardizes cross-system workload identity~\cite{ietf-wimse-arch,wimse-s2s,wimse-credx}.

We present the ICP architecture, use cases, and integration points with cloud-native systems. While we do not provide a full prototype, we include a comparative analysis with existing solutions and theorize performance properties including token validation latency, enforcement scope, and policy drift risks.

\section{Background and Related Work}

\subsection{Workload Identity and SPIFFE}
The Secure Production Identity Framework for Everyone (SPIFFE) standard~\cite{spiffe} enables workload-to-workload authentication using X.509 or JWT-based service identity documents (SVIDs). SPIRE, its reference implementation, provides workload attestation and trust domain management. SPIFFE decouples identity from infrastructure topology and supports service mesh enforcement (e.g., Istio), but does not unify identity across humans and automation.

\subsection{Human Identity: OIDC and SAML}
OpenID Connect (OIDC) and SAML are widely adopted for user authentication and federated access. While effective for SSO and role assignment, these models lack ephemeral, intent-bound identity constructs and are unsuitable for runtime authorization or scoped policy enforcement in automation and infrastructure workflows.

\subsection{IETF WIMSE: Workload Identity in Multi-System Environments}
The IETF WIMSE working group defines standards for federated identity and credential exchange across systems~\cite{ietf-wimse-arch}. Its \textit{Workload Identity Token} and \textit{Credential Exchange} drafts~\cite{wimse-s2s,wimse-credx} provide a foundation for secure, context-rich authentication between actors operating in distinct trust domains. The Identity Control Plane adopts and operationalizes these principles.

\subsection{ABAC and Policy Engines}
Attribute-based access control (ABAC) has gained adoption through engines like Open Policy Agent (OPA)~\cite{opa} and Cedar~\cite{cedar}. These systems evaluate policies using actor metadata, environment context, and resource tags. However, they require a consistent identity substrate and integration with runtime execution layers to enforce policies reliably.

\subsection{Credential Brokers and Automation Identity}
Credential brokers issue scoped, short-lived tokens for automation tasks, typically through external identity verification and context evaluation. Prior work on Zero Trust CI/CD~\cite{ci_cd_id} and credential delegation~\cite{broker} highlighted the need for ephemeral, intent-scoped automation identity in pipelines and ephemeral environments.

\subsection{Related Architectures and Gaps}
\begin{table}[h]
\caption{Comparison of Identity Enforcement Models}
\centering
\resizebox{\columnwidth}{!}{%
\begin{tabular}{|l|c|c|c|c|c|}
\hline
\textbf{Model} & \textbf{Scope} & \textbf{Federation} & \textbf{Tokenization} & \textbf{Policy} & \textbf{Enforcement} \\
\hline
AWS IAM & Workload/User & Limited & Static & RBAC & Cloud API \\
Istio + SPIFFE & Workload & Yes & SVID & RBAC & Proxy \\
GitHub OIDC & User/Automation & Medium & OIDC JWT & Limited & CI/CD \\
\textbf{ICP (This Work)} & All & Full & Scoped & ABAC & Multi-Layer \\
\hline
\end{tabular}%
}
\end{table}

As shown in Table I, existing systems either focus on one actor type or lack runtime enforcement. The Identity Control Plane fills this gap by offering a composable framework for real-time, policy-driven identity orchestration across diverse control planes.

\section{Problem Definition}
Cloud-native infrastructure operates across multiple trust domains, identity sources, and operational planes. Despite widespread adoption of Zero Trust principles, identity systems remain disjointed and inconsistent in their enforcement capabilities.

\subsection{Fragmented Identity Models}
Human, workload, and automation identities are issued and verified by different systems—OIDC for users, SPIFFE or cloud IAM for services, and shared secrets for automation. These silos inhibit secure delegation and increase operational complexity.

\subsection{Lack of Intent-Bound Credentials}
Automation workflows often rely on long-lived service accounts or static role bindings. These credentials do not encode transaction context (e.g., commit SHA, actor, purpose), making fine-grained enforcement and post-hoc attribution difficult.

\subsection{Static Policy Models}
RBAC and IAM policies are typically static and environment-agnostic. They lack support for runtime metadata such as deployment state, environment tags, or triggering pipeline. This leads to over-permissioning and brittle policy definitions.

\subsection{No Unified Control Layer}
There is no central mechanism to normalize identity across systems and enforce attribute-based policies at the point of execution. Existing IDPs provide authentication but not identity-bound enforcement or simulation at the infrastructure layer.

\subsection{Audit and Compliance Blind Spots}
Without a shared identity substrate, access logs are fragmented. Most systems cannot map resource access back to the initiating identity in a verifiable, cryptographically secure way. This inhibits auditability and regulatory compliance.

\textbf{Summary:} A scalable Zero Trust model requires a unified, identity-aware control plane that enables scoped, intent-driven authorization and enforcement across users, workloads, and automation.

\section{Identity Control Plane Architecture}
The Identity Control Plane (ICP) unifies authentication, tokenization, and policy enforcement for all identity types—human, workload, and automation. It abstracts identities across trust domains and evaluates access dynamically using attribute-based policies.

\subsection{Core Components}
\textbf{1) Identity Abstraction Layer:}
Normalizes identity formats using:
\begin{itemize}
    \item SPIFFE IDs for workloads~\cite{spiffe}
    \item OIDC/SAML for human users
    \item Broker-issued transaction tokens for automation and CI/CD jobs
\end{itemize}

\textbf{2) Credential Brokers:}
Generate scoped, short-lived tokens. Tokens embed metadata including actor, source system, environment, time, and purpose.

\textbf{3) Policy Evaluation Layer:}
Enforces access using ABAC engines like OPA~\cite{opa} and Cedar~\cite{cedar}. Policies reference identity, resource metadata, and real-time intent (e.g., environment, git ref, action type).

\textbf{4) Metadata Integration:}
Pulls context from internal developer platforms (IDPs), source control, Kubernetes metadata, and cloud control planes.

\subsection{Modes of Operation}
\begin{itemize}
    \item \textbf{Real-Time Enforcement:} Policy evaluated at the moment of action (e.g., API request, deployment, shell access).
    \item \textbf{Intent Evaluation:} Dry-run simulation to validate access before execution (e.g., PR checks, deployment preview).
    \item \textbf{Replay and Audit:} Post-event simulation to explain policy behavior or validate policy changes retroactively.
\end{itemize}

\subsection{SPIFFE Bundle Synchronization for Cross-Domain Trust}
To enable secure cross-domain workload authentication, SPIFFE uses trust bundles that define the root CA certificates for other trust domains~\cite{spiffe}. The Identity Control Plane incorporates bundle synchronization mechanisms—either via SPIRE federation APIs or through automated policy distribution—to establish mutual trust between disparate SPIFFE domains. This allows for authenticated service-to-service communication between different cloud accounts, Kubernetes clusters, or organizational boundaries without requiring shared identity providers.

\subsection{Reference Architecture Diagram (Textual)}
\begin{figure}[h]
\centering
\includegraphics[width=0.45\textwidth]{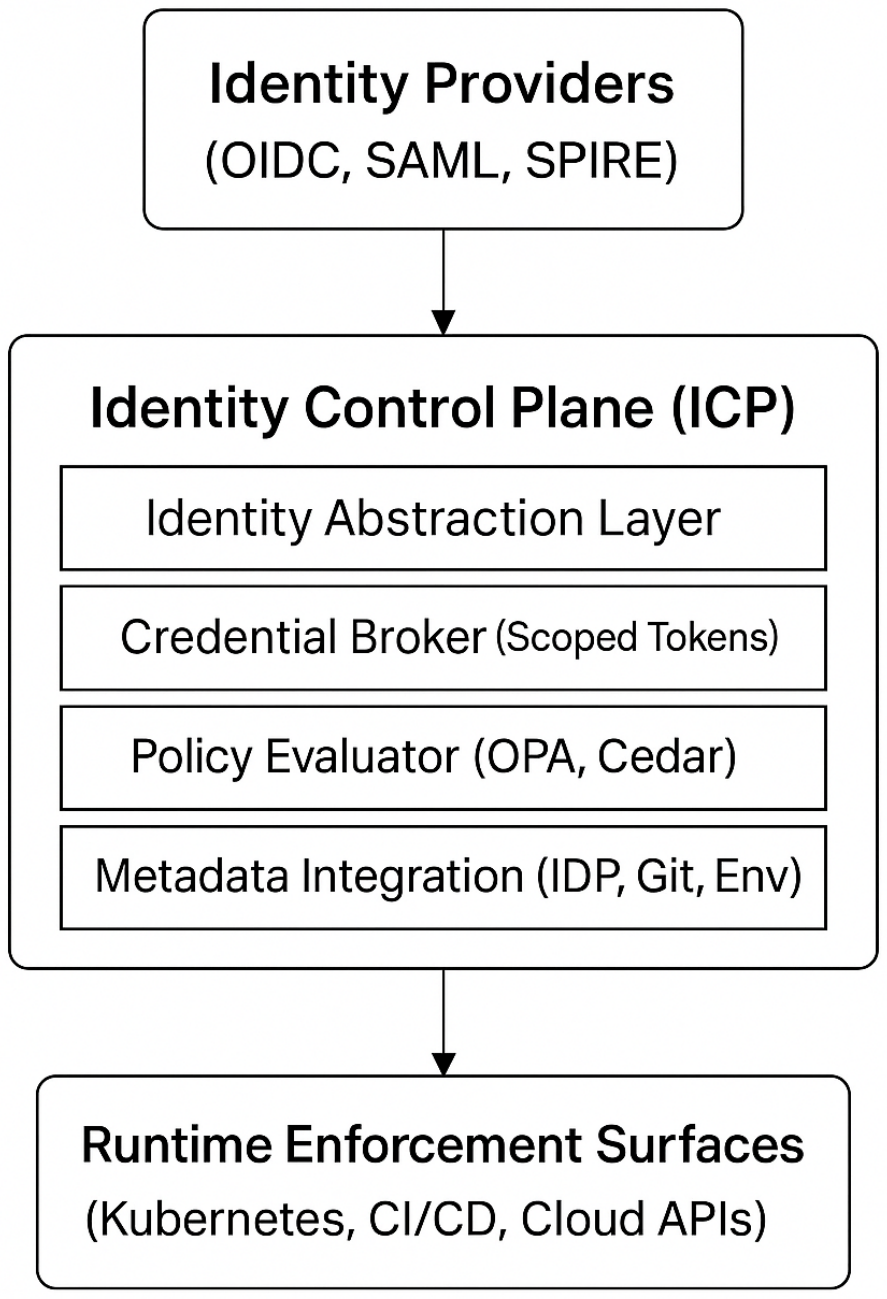}
\caption{Identity Control Plane Architecture (Reference)}
\label{fig:icp-arch}
\end{figure}

The ICP integrates into:
\begin{itemize}
    \item CI/CD job runners
    \item Service mesh proxies (e.g., Envoy with SPIFFE)
    \item Kubernetes admission controllers
    \item Cloud API gateways
    \item Developer portals (for just-in-time access requests)
\end{itemize}

The architecture enables organizations to enforce Zero Trust policies across heterogeneous systems using a composable, scalable identity control framework.

\section{Use Cases}
The Identity Control Plane (ICP) enables secure, context-aware identity enforcement across multiple operational scenarios. These use cases demonstrate how ICP supports Zero Trust in practice.

\subsection{Platform Onboarding}
When a new service is onboarded, the ICP assigns a SPIFFE ID and binds contextual metadata (e.g., team, environment, application tag). Developers receive scoped tokens based on ownership and intent. ABAC policies ensure onboarding actions conform to organizational guardrails.

\subsection{CI/CD Deployments}
Each pipeline execution receives an ephemeral identity and brokered transaction token. The token includes metadata such as commit hash, branch name, and pipeline actor. Policies restrict deployment to authorized environments and enforce provenance and approval checks.

\subsection{Developer Ephemeral Access}
Developers request temporary access to environments (e.g., staging, debug pods) through a portal integrated with the ICP. The system evaluates policy based on identity, request context, and environment. If permitted, the developer receives a time-bound token with observability and automatic revocation.

\subsection{Cross-Account Access Delegation}
For multi-cloud and multi-account architectures, ICP federates identity using SPIFFE Trust Domains and OIDC chains. A credential broker issues a scoped token valid in the target domain, validated against policies that account for originating identity, resource scope, and intent.

\subsection{Legacy System Integration}
ICP supports legacy systems that lack native SPIFFE/OIDC support through sidecar proxies or gateway enforcement. These components validate tokens, enforce ABAC policies, and inject scoped credentials into legacy systems while maintaining a unified audit trail.

\subsection{Policy Simulation for Preflight Enforcement}
Before merging infrastructure-as-code or triggering deployment workflows, developers can simulate ICP policy decisions using a dry-run API. This improves predictability, reduces rollout errors, and supports shift-left security practices.

\section{Evaluation and Integration Patterns}
The Identity Control Plane (ICP) is designed to integrate with existing infrastructure systems while enforcing dynamic, intent-aware policies at runtime.

\subsection{Interoperability with Identity Systems}
ICP supports identity ingestion and federation from:
\begin{itemize}
    \item SPIFFE/SPIRE for workload identity~\cite{spiffe}
    \item OIDC/SAML for user identity
    \item GitHub Actions, GitLab, and CI/CD providers for automation identity
\end{itemize}
It bridges identity silos by normalizing identities into a unified, policy-enforceable format compatible with runtime systems and policy engines.

\subsection{Integration Points}
ICP operates across multiple control surfaces:
\begin{itemize}
    \item CI/CD Runners: Inject scoped tokens during job execution
    \item Service Mesh Proxies: Enforce mutual TLS and SPIFFE identity checks
    \item Kubernetes Admission Controllers: Bind metadata and enforce identity-aware policies
    \item Infrastructure APIs and Gateways: Validate access requests using transaction tokens
    \item Developer Portals: Evaluate and simulate policies preflight
\end{itemize}

\subsection{Theorized Performance Characteristics}
\begin{itemize}
    \item Token issuance latency: 50–100 ms via local broker with identity attestation
    \item Policy evaluation time: \(<10\ \mathrm{ms}\) per request
 OPA~\cite{opa} or Cedar~\cite{cedar}
    \item Token validation overhead: Negligible for stateless JWT-based enforcement
    \item Control plane scalability: Stateless evaluation enables distributed enforcement
\end{itemize}

\subsection{Security and Operational Benefits}
\begin{itemize}
    \item Reduced Blast Radius: Scoped, short-lived tokens minimize lateral movement
    \item Unified Policy Enforcement: ABAC policies apply uniformly across identity types
    \item Improved Observability: All access is verifiable, traceable, and auditable
    \item Compliance Alignment: Supports SLSA, SOC 2, FedRAMP, and other runtime evaluation standards~\cite{compliance}
    \item Policy Simulation: Enables dry-run evaluations and intent validation prior to execution
\end{itemize}

\subsection{Limitations and Trade-Offs}
\begin{itemize}
    \item Policy Complexity: Requires robust modeling of identity and metadata
    \item Latency Spikes: Broker or policy engine bottlenecks may delay enforcement
    \item Federation Risk: Poorly scoped trust domains or token misuse may weaken boundaries
    \item Legacy System Compatibility: May require proxies or enforcement shims for integration
\end{itemize}
Despite these trade-offs, ICP significantly improves identity-driven access enforcement across modern infrastructure, enabling a scalable foundation for Zero Trust.

\section{Discussion and Future Work}
\subsection{Policy-as-Intent Evolution}
The ICP aligns with the industry shift from static RBAC to intent-aware, dynamic authorization. Policies are evaluated not just on identity, but also on purpose, scope, and metadata. This allows security teams to reason about ``why'' an action occurred, not just ``who'' performed it.

\subsection{Compliance and Continuous Validation}
ICP enables continuous compliance by embedding policy checks into infrastructure workflows. Pre-deployment simulation, real-time evaluation, and post-hoc replay collectively support traceable, auditable control enforcement. This model aligns with compliance frameworks such as SOC 2, FedRAMP, and SLSA.

\subsection{Agentic Workflows and AI Integration}
As intelligent agents increasingly perform operational tasks (e.g., LLM-driven DevOps assistants), ICP can assign identity-scoped, session-bound tokens. This enables auditability and revocation of AI-generated actions. Future work may explore zero-knowledge delegation or fine-grained proof of intent models.

\subsection{Adoption Roadmap}
Enterprises can adopt ICP incrementally:
\begin{enumerate}
    \item Introduce SPIFFE-based identity for workloads and CI/CD runners
    \item Deploy credential brokers to issue scoped transaction tokens
    \item Migrate static IAM policies to ABAC engines (OPA~\cite{opa}/Cedar~\cite{cedar})
    \item Integrate policy simulation into developer workflows
    \item Extend enforcement to cross-account and human access
\end{enumerate}

\subsection{Security and Ethical Considerations}
ICP assumes trusted identity issuance, well-scoped policies, and secure token handling. Risks include:
\begin{itemize}
    \item Federation drift: Misaligned trust boundaries may lead to privilege escalation
    \item Over-delegation: Improper token scopes could enable unintended access
    \item Blind spots in simulation: Incomplete metadata or assumptions may bypass controls
\end{itemize}

\subsection{Compliance Case Study: FedRAMP and SLSA Alignment}
The ICP architecture directly supports runtime controls mandated by compliance frameworks such as FedRAMP and Supply Chain Levels for Software Artifacts (SLSA). For example:
\begin{itemize}
    \item SLSA Provenance: Scoped tokens issued at CI/CD runtime can encode git metadata, commit SHA, and signer identity—supporting SLSA Level 2+ provenance.
    \item FedRAMP Controls: ICP supports real-time ABAC enforcement (AC-2, AC-3), token revocation (AC-12), and access audits (AU-2, AU-6)~\cite{compliance}. These capabilities can be mapped to existing OSCAL-based control catalogs for formal audit integration.
\end{itemize}
Future research should explore policy verifiability, trust minimization in token issuance, and transparent audit tooling to support ethical use.

\section{Conclusion}
Modern cloud environments require secure, intent-aware access control across users, workloads, and automation systems. Existing identity models are fragmented and unable to enforce unified policies across trust domains, infrastructure boundaries, and operational contexts.

This paper introduces the Identity Control Plane (ICP), an architectural model that integrates workload identity (via SPIFFE~\cite{spiffe}), human identity (via OIDC/SAML), and scoped automation credentials (via transaction tokens) into a single, composable enforcement layer. Policies are enforced using ABAC engines~\cite{opa,cedar} with context from developer platforms, cloud control planes, and CI/CD metadata.

The ICP model enables real-time enforcement, preflight simulation, and post-action audit replay. It aligns with the IETF WIMSE initiative~\cite{ietf-wimse-arch} and OAuth token standards, operationalizing workload identity federation and cross-system credential exchange at scale.

We detail its architecture, use cases, integration points, and performance characteristics. While not prototyped, we theorize its feasibility based on components in production use today.

By shifting identity from a prerequisite to a runtime control surface, the Identity Control Plane provides a foundation for scalable, verifiable, Zero Trust infrastructure.

\textbf{Identity is no longer the perimeter—it is the control plane.}

\end{document}